\documentclass[a4paper,12pt,english,german]{article}

\usepackage{caption}
\usepackage{subcaption}

\usepackage{mathtools}
\usepackage{graphicx}

\begin{document}

{\bf Dynamical stability of the one-dimensional rigid Brownian rotator: The role of the rotator's spatial size and shape}

\bigskip

{Jasmina Jekni\' c-Dugi\' c$^a$, Igor Petrovi\' c$^b$, Momir Arsenijevi\'
c$^c$, Miroljub Dugi\' c$^{\ast c}$

\bigskip

$^a$University of Ni\v s, Faculty of Science and Mathematics, Vi\v
segradska 33, 18000 Ni\v s, Serbia

$^b$Svetog Save 98, 18230 Sokobanja, Serbia

$^c$University of Kragujevac, Faculty of Science, Radoja
Domanovi\' ca 12, 34000 Kragujevac, Serbia

\bigskip

{\bf Abstract} We investigate dynamical stability of a single propeller-like shaped molecular cogwheel modelled as the fixed-axis rigid rotator. In the realistic situations,
rotation of the finite-size cogwheel is subject of the envi\-ron\-men\-tally-induced Brownian-motion effect that we describe by utilizing the quantum Caldeira-Leggett
master equation. Assuming the initially narrow (classical-like) standard deviations for the angle and the angular momentum of the rotator, we investigate
dynamics of the first and second moments depending on the size, i.e., on the number of blades of both the free rotator as well as of the rotator in the external harmonic field.
The larger the standard deviations, the less stable (i.e. less predictable) rotation.
We detect the absence of the simple and straightforward rules for utilizing the rotator's stability. Instead, a number of the size-related criteria appear whose combinations may provide the optimal rules for the rotator dynamical stability and possibly control. In the realistic situations, the quantum-mechanical corrections, albeit individually small, may effectively prove non-negligible, and also revealing
subtlety of the transition from the quantum to the classical dynamics of the rotator.
As to the latter, we detect a strong size-dependence of the transition to the classical dynamics beyond the quantum decoherence process.

\bigskip

{\bf 1. Introduction}

\bigskip

The standard theory of the translational Brownian motion can be formulated without making any reference to the Brownian particle's spatial shape and size. Often, the Brownian particle is imagined as a material point with the phenomenologically modeled external noise and friction [1-3]. Interestingly enough, this extremely idealized model has a remarkable history of success regarding various tasks in physics, chemistry and applications.

	Introducing the Brownian-particle's size typically leads to considerations of spherical/ellipsoid objects, in which case the rotational motion also must be accounted, see e.g. [4, 5].  As a geometrical limit of the ellipsoid-like Brownian particle, most often is regarded a rod-like rotator [5] (and references therein). Those idealized geometrical models  may be useful for the sufficiently large particles--of the diameter of the order of $100$nm. However, for the smaller particles, these models may not be reliable anymore.

	The functional parts of the desired nano-machines, such as the molecular nano-rotators (“molecular cogwheels”) [6-9], are purposefully designed--their function strongly depends on their geometrical shape and size, which is typically of the order of $1-10$nm. These spatial dimensions are such that the molecule geometrical shapes can be “seen” by the environmental particles so much that the standard approximations of the Brownian particle, e.g. by the spherical-like models, may not in general be viable. Furthermore, when immersed in a solution or even resting on surfaces (as it might be in practice), significant thermally driven, i.e. random, molecular rotations may be expected. Interestingly enough, in some cases, thermal noise can assist directed rotation [10], thus revealing subtleties, and possibly unexpected falls of our classical intuition regarding the spatially structured microscopic systems. That is, macroscopic analogies do not necessarily go far in predicting function in nanoscale environments. The analogies may break due to the environmental influence (such as Brownian motion) or various kinds of quantum effects or both [11, 12].

	Following the existing candidates for the really geared and possibly controllable molecular nano-rotators, in this paper, we consider the {\it propeller-like shaped} molecular species [6-9] that we model as  Brownian rotators. For simplicity, we assume one-dimensional (fixed axis) rigid rotator. Spatial {\it size of the rotator} is introduced by the number $N$ of the blades, where the average moment of inertia, denoted $I_{\circ}$, and the average damping rate, denoted $\gamma_{\circ}$, serve as the physical units for the model. In order to tackle the possibly realistic situations [6-9], we consider the number of blades $1\le N \le 10$.

	Our objective is to investigate the rotational stability, i.e.  predictability of rotation (and therefore rotation controllability), depending on the rotator's size and shape. Quantitatively, we investigate the dynamics of the first and second moments for a free and “harmonic” rotator modeled by the standard Caldeira-Leggett (CL) master equation [13, 14] and follow the standard rule [15]: the smaller the standard deviations, the more
stable (i.e. more predictable) dynamics.

The use of the CL master equation is motivated by both, the well-defined classical limit as well as by the explicit quantum-mechanical corrections in the {\it weak}-coupling limit. Therefore our main focus is placed on both the underdamped and non-underdamped regimes [below precisely to be defined]; extensions of these considerations are discussed in Section 4.

	We find the absence of the straightforward recipes for utilizing the relative stability of the rotation. The choice of the {\it optimal conditions} for the desired control of the rotation depends on a number of criteria, notably of: (i) the standard deviations of the angle and the angular momentum exhibit the opposite dependencies on the size (the number of blades) of the rotator; (ii) the significantly different magnitudes of the change for the angle- and the angular-momentum standard deviations; (iii) the shorter relaxation times for larger propellers; (iv) the value of the tunable damping rate $\gamma_{\circ}$; (v) [in  conjunction with the point (iv)] the time scale of the allowed/desired external actions on the rotator. Expectably, {\it quantitatively}, those criteria depend on (a) the presence/absence of the external harmonic field for the rotator as well as on (b) the underdamped/non-underdamped regime. Intuitively, the optimal choice of the conditions for appropriate function of the molecular cogwheels is a reminiscence of the standard engineering optimization [16] already at the microscopic scale. However, here disappears  analogy with the classical counterparts [17].

Our results reveal the possible accumulation and therefore increase of the quantum-mechanical, individually negligible,  effects. The absence of those effects in the classical domain emphasizes a limited use of the classical (or semi-classical) theory of the Brownian rotation [6-9]. Going beyond such treatment is provided in this paper by utilizing the {\it full} Caldeira-Leggett master equation [13, 14].
Hence also the size-dependent transition to the classical dynamics that goes beyond the decoherence process.

In Section 2, we introduce the model of interest, with an emphasis on the subtleties of the transition from the translational to the rotational model. In Section 3 we provide the quantitative results, which are discussed in Section 4 with an emphasis on relaxing the assumptions of the weak coupling and high temperature of the bath. Section 5 is conclusion.

\bigskip

{\bf 2. Quantum Brownian rotator}

\bigskip

The shaded areas in Figure 1 depict the average effective surface for a set of blades exposed to the environmental influence; depending on the blades geometry, the surfaces may be only partly exposed to the environmental influence. The length $L$ and the height $d$ uniquely determine the moment of inertia, denoted $I_{\circ}$, and the strength of interaction, denoted $\alpha_{\circ}$, of the typical blade of a propeller rotating around the fixed axis $z$--in analogy with the, often addressed [17], macroscopic models [18].

\begin{figure}[h]
  \centering
  \includegraphics[width=.5\linewidth]{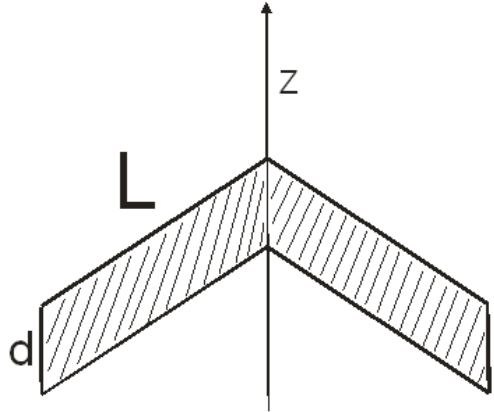}
\caption{A pair of blades illustrating the  average effective size of a set of blades; the shaded areas depict the surfaces that may be exposed to the environmental influence.}
\end{figure}

Both $I_{\circ}$ and $\alpha_{\circ}$} are subjects of free choice in our considerations. On the one hand, the average moment of inertia, $I_{\circ}$, of the typical ("averaged") blade regards the different chemical species and local geometric deformations (e.g. twisted blades) that define the molecules of different (local) configurations, which include the molecules conformations [6]. On the other hand, the strength of interaction can be externally tuned (even by several orders of magnitude) by appropriate choice of the rotator's environment [19] and the only assumption is that the strength of the interaction with the environment
allows for the use of the weak-coupling limit  [14, 20].

From Fig.1 it is obvious that for a propeller with $N$ blades, the total moment of inertia $I=NI_{\circ}$. Now assuming that the strength of the interaction with the environment is proportional to the average effective surface of one blade [14], the strength of interaction for $N$ blades $\alpha = N\alpha_{\circ}$, where $\alpha_{\circ}$ is the average strength of interaction for a set of $N$ blades. Therefore the assumption of the weak coupling limit requires $N_{\max}\alpha_{\circ} = 10\alpha_{\circ} \ll 1$.

Transition from the translational to the rotational Brownian model is formally straightforward [21]: the position operator $ x$ is exchanged by the angle of rotation $\varphi$, momentum $ p$ by the angular momentum $ L_z$, and the mass $m$ by the rotator's moment of inertia $I$. Adopting the basic assumptions of the weak interaction and high temperature in the {\it microscopic} derivation
of the translational model [13, 14] directly gives rise to the Caldeira-Leggett (CL) master equation for the Brownian rotator (in the Schr\" odinger picture) [21]:

\begin{equation}
{d \rho_R(t)\over dt} = -{\imath\over\hbar} [H_R, \rho_R(t)] -
{\imath\gamma\over\hbar} [\varphi, \{L_z, \rho_R(t)\}] - {2I\gamma
k_B T\over \hbar^2}[\varphi, [\varphi, \rho_R(t)]],
\end{equation}

\noindent where $\gamma$ is the damping rate for the rotator and the rotator's self Hamiltonian $H_R=L_z^2/2I+V(\varphi)$.
Eq.(1) has a well-defined classical counterpart in the form of the Langevin equation for the angle variable $\varphi$ [6, 9], in the full formal analogy with the Langevin equation for the Descartes position variable $x$.

The spectral density, denoted
$J$, for the Brownian motion model is proportional to $\alpha^2$
[14] while, on the other hand, $J \propto I\gamma$ (see eq.(3.392)
in [14]).
Then for one blade $\alpha_{\circ}^2\propto I_{\circ}
\gamma_{\circ}$ while for $N$ blades, from $(N\alpha_{\circ})^2
\propto N I_{\circ} \gamma$, we obtain e.g. $\gamma = N\gamma_{\circ}$.
Therefore the linear proportionality of both $I$ and $\gamma$ with the size (with the number $N$ of blades) of the rotator.

We adopt eq.(1) without modifications  with an emphasis on the well-defined classical counterpart. However,
analogy between the translation and rotation breaks in the quantum-mechanical context when it comes to the {\it meaning} of the
standard deviations and the validity of the uncertainty relation for the angle and the angular-momentum observables [22-24].
On the one hand, the standard deviation  $\Delta\varphi = \sqrt{\langle \varphi^2\rangle - \langle\varphi\rangle^2}$ can be used as a measure of uncertainty of the angle $\varphi$ {\it only} for sufficiently
small values [22]. On the other hand, for very small values of $\Delta\varphi$, even for the  finite interval $\varphi\in [0, 2\pi]$, the rotational model becomes formally analogous to the translational model.
Bearing in mind that the uncertainty relation for $\varphi$ and $L_z$ is not well defined, the analogy with the translational motion is lost as we do not use the uncertainty relation as a constraint of our considerations. On the other hand, the use of the standard deviation $\Delta\varphi$ as a measure of uncertainty is provided by assuming the small initial value $\Delta\varphi(0)$.

Now, taking the small initial $\Delta \varphi(0)$ and $\Delta L_z(0)$ makes the whole picture very close to the classical counterpart; in this regard, the details are given below. Here we just want to emphasize that the classical-like rotational dynamics is fully in the spirit of the original CL-dynamics, which assumes the extremely high temperature $T$ of the thermal bath (reservoir)--the underdamped regime.  Nevertheless, wide applicability of the CL master equation [14, 15], i.e. of eq.(1), allows for the considerations in the non-underdamped regime--toward the  more general scenarios discussed in Section 4.

Denoting by $E$ the rotator's self-energy and by $T$ the bath's temperature, the underdamped and the non-underdamped regimes of interest [14, 20] are here defined by the following non-equalities: $N_{\max}\hbar\gamma_{\circ} \ll E < k_BT$ and $N_{\max}\hbar \gamma_{\circ} \ll k_BT < E$, respectively. For the free rotator, $E = \hbar^2/2N_{\max}I_{\circ}$, and for the harmonic rotator, $E= \hbar \omega$; $\omega$ is the frequency of the external harmonic field and $k_B$ is the Boltzmann constant.

In analogy with the standard procedure for the translational
Brownian motion (cf. eqs.(3.426)-(3.430) in Ref. [14]), from eq.(1) easily, and unconditionally, follow the equations for the
first and second moments for the (fixed-axis) rotational Brownian motion:

\begin{eqnarray}
&\nonumber& {d\langle\varphi(t)\rangle\over dt} = {1\over I}
\langle L_z(t)\rangle,\\&& {d\langle L_z(t)\rangle\over dt} = -
\langle V'(\varphi(t))\rangle - 2 \gamma \langle L_z(t)\rangle,
\nonumber
\\&&
{d\langle \varphi^2(t)\rangle\over dt} = {1\over I} \langle
L_z(t)\varphi(t) + \varphi(t) L_z(t) \rangle \nonumber,
\\&& \nonumber
{d\langle L_z^2(t)\rangle\over dt} = - \langle L_z(t)
V'(\varphi(t)) + V'(\varphi(t)) L_z(t)\rangle - 4\gamma \langle
L_z^2(t)\rangle + 4I\gamma k_B T
\\&& {d\over dt}\langle \varphi L_z + L_z\varphi \rangle = {2\over
I}\langle L_z^2\rangle -2\langle \varphi V'(\varphi)\rangle -
2\gamma \langle \varphi L_z + L_z\varphi \rangle.
\end{eqnarray}

\noindent In eq.(2), $V'(\varphi(t))\equiv
dV(\varphi(t))/d\varphi(t)$, while the mean values $\langle
\ast\rangle = tr (\ast \rho_R(t))$.
Eq.(2) is general--it applies for every kind of the external field $V(\varphi)$. In this paper we will consider only the free rotator ($V=0$) and the rotator in the external harmonic field ($V(\varphi) = I\omega^2\varphi^2/2$) for small values of the angle of rotation. A proper series of small rotations can effect in the finite rotation of the molecule [6-8].

Our task in this and the next section is to investigate dependence on the number $N$ of blades of the moments in eq.(2) for both, the free and harmonic rotator, in the underdamped and non-underdamped regimes.
The only constraints come from the assumptions of the very small initial $\Delta\varphi$ and small angle of rotation as well as from the requirement of the weak-coupling and high temperature limit [14, 20].

\bigskip

{\bf 2.1 Free rotator}

\bigskip

For the free rotator, $V=0$, the solutions of eq.(2) can be directly taken over from the solutions for the free translational motion, eqs.(3.438)-(3.440) in Ref. [14], from which easily follows also the solution for the variance function, $\sigma_{\varphi L} = \langle \varphi L_z + L_z\varphi\rangle - 2\langle \varphi\rangle \langle L_z\rangle$, which in the classical limit takes the form $\sigma_{\varphi L} = \langle \varphi L_z\rangle - \langle\varphi\rangle \langle L_z\rangle$. Dividing by $I_{\circ} \gamma_{\circ}$, we obtain the dimensionless quantities $\sigma_L \equiv \Delta L_z/I_{\circ}\gamma_{\circ}$, $\sigma\equiv \sigma_{\varphi L}/I_{\circ}\gamma_{\circ}$ and   $\tau \equiv \gamma_{\circ} t$:

\begin{eqnarray}
&\nonumber& \sigma_{\varphi}^2(\tau) = \sigma_{\varphi}^2(0) +
\left({1 - e^{-2N \tau}\over 2N^2}\right)^2 \sigma_L^2(0) + {1 -
e^{-2N \tau}\over 2N^2} \sigma(0) \\&& \nonumber + {k_B T \over
N^3 I_{\circ}\gamma_{\circ}^2} \left[N \tau - (1 - e^{-2N \tau}) +
{1\over 4}(1 - e^{-4N \tau})\right],
\\&& \nonumber \sigma_L^2(\tau) = e^{-4N \tau} \sigma_L^2(0) + N {k_B T \over
I_{\circ}\gamma_{\circ}^2} (1 - e^{-4N \tau}),
\\&&
\sigma(\tau) = \sigma(0) e^{-2N\tau} + {e^{-2N\tau}\over N^2}(1-e^{-2N\tau}) \sigma_L^2(0) + {k_BT\over N I_{\circ}\gamma_{\circ}^2} (1-e^{-2N\tau})^2.
\end{eqnarray}

\noindent In the asymptotic limit ($\tau\to\infty$), from eq.(3):

\begin{equation}
\lim_{\tau\to\infty}\sigma_{\varphi} = \sqrt{ {k_B T \over
I_{\circ} \gamma_{\circ}^2} {\tau\over N^2} }, \quad
\lim_{\tau\to\infty} \sigma_L = \sqrt{ {N k_B T \over I_{\circ}
\gamma_{\circ}^2} }, \quad \lim_{\tau\to\infty}\sigma = {k_BT\over N I_{\circ}\gamma_{\circ}^2}.
\end{equation}

The classical counterparts of eq.(3) are given in Appendix I. The classical variance is identical with the quantum-mechanical expression, while the classical expressions for $\sigma_L$ and $\sigma_{\varphi}$ follow from  placing $\sigma_L(0) = \sigma_{\varphi}(0) = \sigma(0) = 0$ in eq.(3). Therefore the clear distinction between the classical and quantum-mechanical contributions in eq.(3).

\bigskip

{\bf 2.2 Harmonic rotator}

\bigskip

Solutions to eq.(2) for the harmonic rotator are obtained in the full analogy with Section 2.1, for the case $V(\varphi)=I\omega^2\varphi^2/2$. To this end, we borrow the solutions from [25], again presented via the dimensionless quantities:

\begin{eqnarray}
&\nonumber& \sigma_{\varphi}^2(\tau) = e^{-2N\tau} \left(
\sigma_{\varphi}^2(0)\cos^2{\omega\over\gamma_{\circ}}\tau +
{\sigma(0)\over 2N}\sin2{\omega\over\gamma_{\circ}}\tau +
{\sigma_L^2(0)\over N^2}\sin^2{\omega\over\gamma_{\circ}}\tau
\right)
\\&&\nonumber + {k_B T\over NI_{\circ}\omega^2}(1 - e^{-2N\tau}) +
O\left({\gamma\over\omega}\right),
\\&& \nonumber
\sigma_L^2(\tau) = e^{-2N\tau} \left( \sigma_L^2(0)
\cos^2{\omega\over\gamma_{\circ}}\tau + N^2\sigma_{\varphi}^2(0)
\sin^2 {\omega\over\gamma_{\circ}}\tau - {N \over 2} \sigma(0)
\sin 2{\omega\over\gamma_{\circ}}\tau\right)
\\&&\nonumber + {N k_B T\over I_{\circ}\omega^2} (1 - e^{-2N\tau}) +
O\left({\gamma\over\omega}\right)
\\&& \nonumber \sigma(\tau) = e^{-2N\tau}
\left(4N {\gamma_{\circ}k_BT\over I_{\circ}\omega^3} \sin^2{\omega\tau\over\gamma_{\circ}}
- N\sigma_{\varphi}^2(0)\sin{2\omega\tau\over\gamma_{\circ}}
+ \sigma(0) \cos{2\omega\tau\over\gamma_{\circ}} +
{\sigma_L^2(0)\over N}\sin{2\omega\tau\over\gamma_{\circ}}\right)\\&& + O\left({\gamma\over\omega}\right).
\end{eqnarray}

\noindent In the asymptotic limit ($\tau\to\infty$), from eq.(5)
with neglecting the terms proportional to $\gamma/\omega$:

\begin{equation}
\lim_{\tau\to\infty}\sigma_{\varphi} = \sqrt{ k_B T\over
NI_{\circ}\omega^2 },\quad \lim_{\tau\to\infty} \sigma_L = \sqrt{ {N
k_B T \over I_{\circ} \omega^2}},\quad \lim_{\tau\to\infty}\sigma = 0.
\end{equation}

\noindent Eq.(6) reduces to eq.(3.424) in [14] that follows from
an approximate treatment of the one-dimensional translational Brownian
particle in the external harmonic field.

The classical expressions are shown in Appendix I to follow from eq.(5) for $\sigma_L(0) = \sigma_{\varphi}(0) = \sigma(0) = 0$. Therefore the clear distinction between the classical and quantum-mechanical contributions in eq.(5).

\bigskip

{\bf 3. Results}

\bigskip

In this section, we utilize  equations (3) and (5) to investigate the rotator's stability as a function of the rotator's size (i.e. of the number $N$ of the blades), and also
 directly compare the exact quantum-mechanical expressions with their classical counterparts.
Of course, the larger the standard deviations the less stability (and hence the less control) of the system.

The results presented below are givenfor the initial values $\sigma_{\varphi}= \sigma = 0.1 \ll 2\pi$ and $\sigma_{ L} = 1 $. Bearing in mind that the typical values of the moment of inertia
$I_{\circ}\sim 10^{-46}-10^{-44}$, for the choice $\gamma_{\circ}< 10^8$ [in the SI units], $\Delta L_z < 10^{-2}\hbar$. Hence $\Delta\varphi \cdot \Delta L_z < 10^{-3}\hbar$, which is the very classical initial condition,
in accordance with ignoring the uncertainty relations for the angle and the angular momentum observables (cf. Section 2).

In eq.(3) appears the constant $A\equiv k_BT/I_{\circ}\gamma_{\circ}^2$, while in eq.(5) appear the constants $B\equiv k_BT/I_{\circ}\omega^2$ and $C = \gamma_{\circ}B/\omega$. The weak-coupling limit allows for the choice
$\gamma_{\circ}/\omega \sim 10^{-3}$. For the free rotator, whose self-energy $H=L_z^2/2I$, we obtain the same value of the $A$ for both regimes:

\begin{equation}
A =   {k_BT \over  \hbar \gamma_{\circ}} \cdot {\hbar \over I_{\circ}\gamma_{\circ}}
\end{equation}

\noindent while $k_BT/\hbar \gamma_{\circ}\gg 1$ and $\hbar/I_{\circ}\gamma_{\circ}\gg 200$. Without loss of generality,  we choose the value $A = 2\cdot 10^5$.

The values of $B$ and $C$ depend on the regime--underdamped or non-underdamped regime for the rotator. Similar to the free rotator, we use:

\begin{equation}
B = {k_BT\over \hbar \omega} \cdot {\hbar\over I_{\circ}\omega}.
\end{equation}

\noindent The constant $k_BT/\hbar\omega >1$ for the underdamped, and  $k_BT/\hbar\omega <1$ for the non-underdamped regime.
Now, assuming that
$\hbar \omega \sim\langle H\rangle = \langle L_z^2\rangle/2I_{\circ} + I_{\circ}\omega^2\langle \varphi^2\rangle/2 \sim I_{\circ}\omega^2$, i.e. that $\hbar/ I_{\circ}\omega \sim 1$ , we take for the underdamped regime, $B \sim 10, C \sim 10^{-2}$, and for the non-underdamped regime, $B \sim 0.1, C \sim 10^{-4}$.

The characteristic time scale of $(N\gamma_{\circ})^{-1}$ is the "relaxation time", for which  the rotator's dynamics becomes essentially classical, i.e. properly described by the  classical Langevin equation for the angle of rotation $\varphi$ [6] that is presented in Appendix I. Nevertheless, in this section we utilize the advantage of the full Caldeira-Leggett model in that we can describe dynamics also for shorter time
intervals. Physically, "shorter"/"longer" time intervals in the units of $(N\gamma_{\circ})^{-1}$ is a {\it relative} concept--as we emphasize in Section 4, it depends on the desired and available operations on the rotator system
so much that even e.g. $t \sim(N\gamma_{\circ})^{-4}$ may not describe the physically noninteresting "transient" processes. For this reason, below, we present the results for the time scale $\tau = \gamma_{\circ} t = 0.01$.

With the use of those values for the parameters, we investigate the equations (3) and (5) and obtain the results as presented in the rest of this section.
Qualitatively, the choice of the initial values as well as the values of the parameters do not change the results to be presented below.
The {\it general} findings are as follows. For every size of the rotator (the number of blades $N$), we observe relatively fast transition to the classical dynamics. For certain shorter time intervals, quantum corrections to the purely classical dynamics are easily detectable--the "quantum domain" of the rotators behavior that is discussed in Section 4--and relatively quickly decrease to reach the classically predicted values with the common asymptotic values for the moments.

In all figures, the surfaces presenting the classical cases are below the quantum-mechanical counterparts--the quantum corrections make the dynamics less stable.

\bigskip

{\bf 3.1 Underdamped regime}

\bigskip

\begin{figure}[h]
\begin{subfigure}{.5\textwidth}
  \centering
  \includegraphics[width=.8\linewidth]{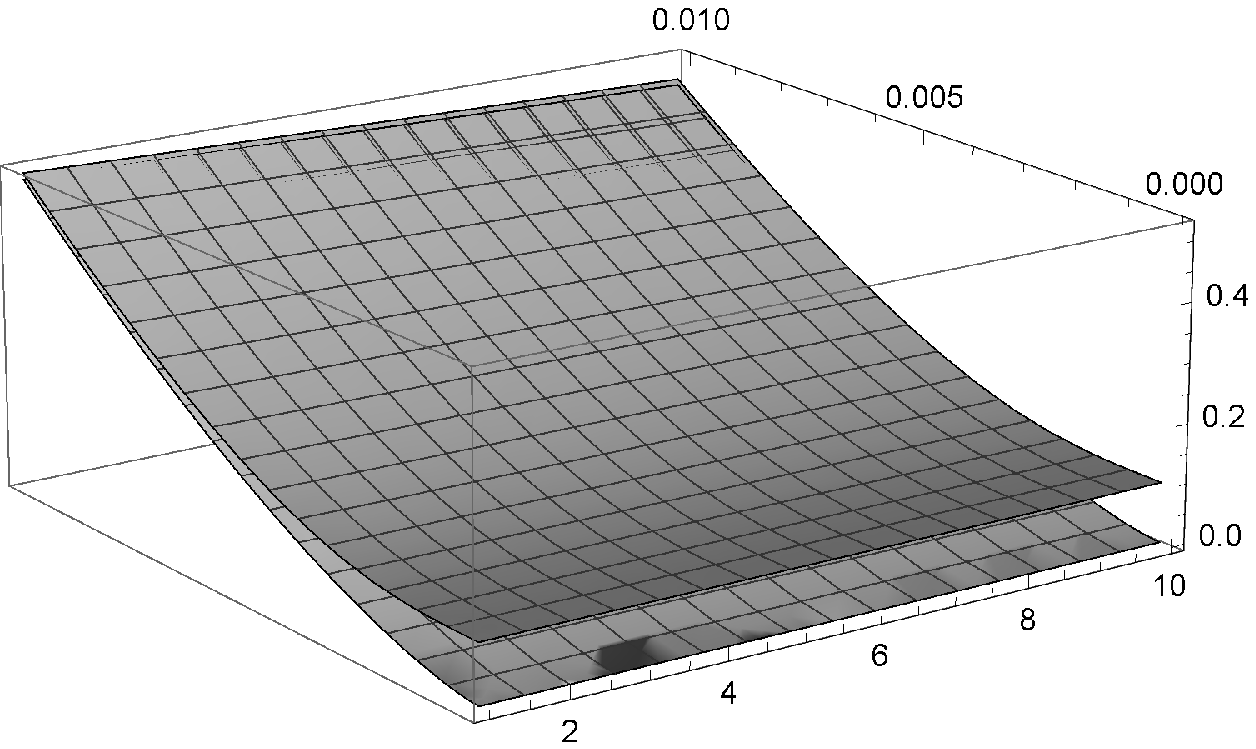}
  \caption{}
\end{subfigure}%
\begin{subfigure}{.5\textwidth}
  \centering
  \includegraphics[width=.8\linewidth]{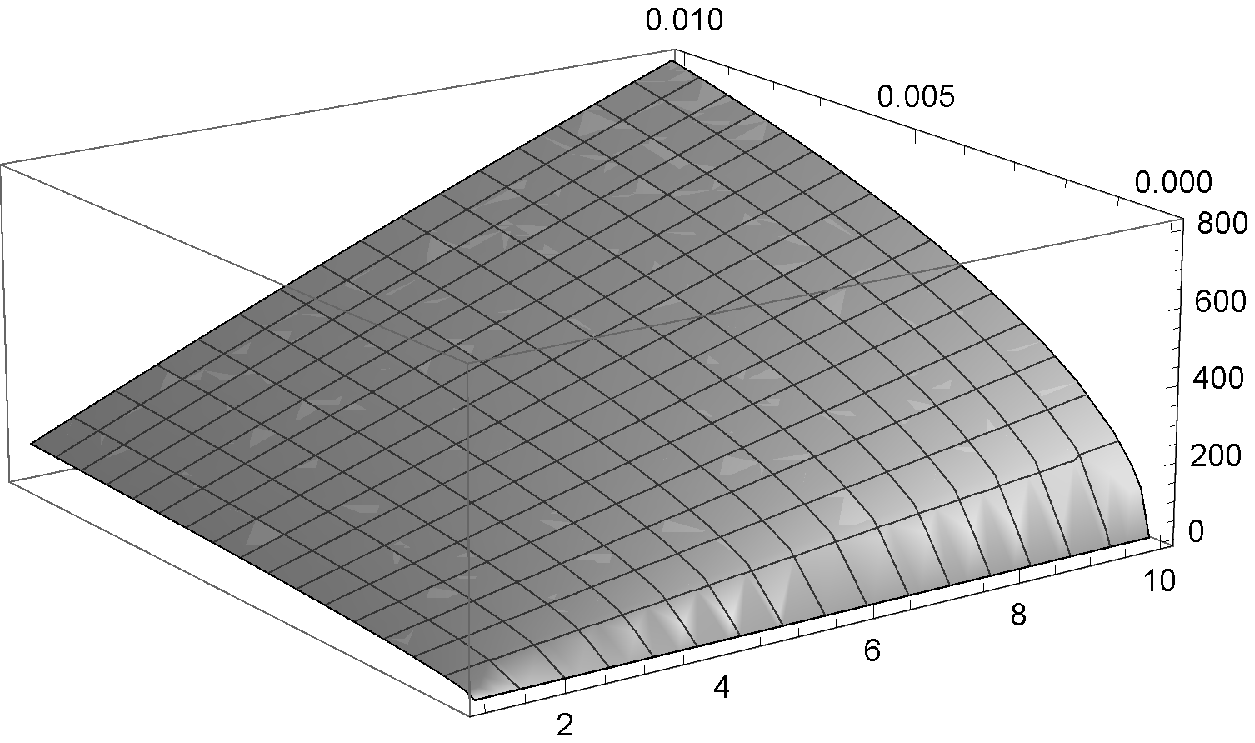}
  \caption{}
\end{subfigure}
\caption{Free rotator, underdamped regime: (a) angle, (b) angular momentum, where for the chosen time scale, the quantum and classical plots are very close to each other. $\tau\in[0, 0.01]$, $N\in[1, 10]$.} \label{Fig1}
\end{figure}

From Figure 2 we can learn about fast transition to the classical dynamics for both $\varphi$ and $L_z$.
Regarding the angle, we find the larger propellers more stable than the smaller ones--in contrast to the angular momentum. The magnitudes of change
of the initial values are remarkably larger for the angular momentum, for which also a significantly faster increase of the standard deviation appears for larger propellers.

\begin{figure}[h]
\begin{subfigure}{.5\textwidth}
  \centering
  \includegraphics[width=.8\linewidth]{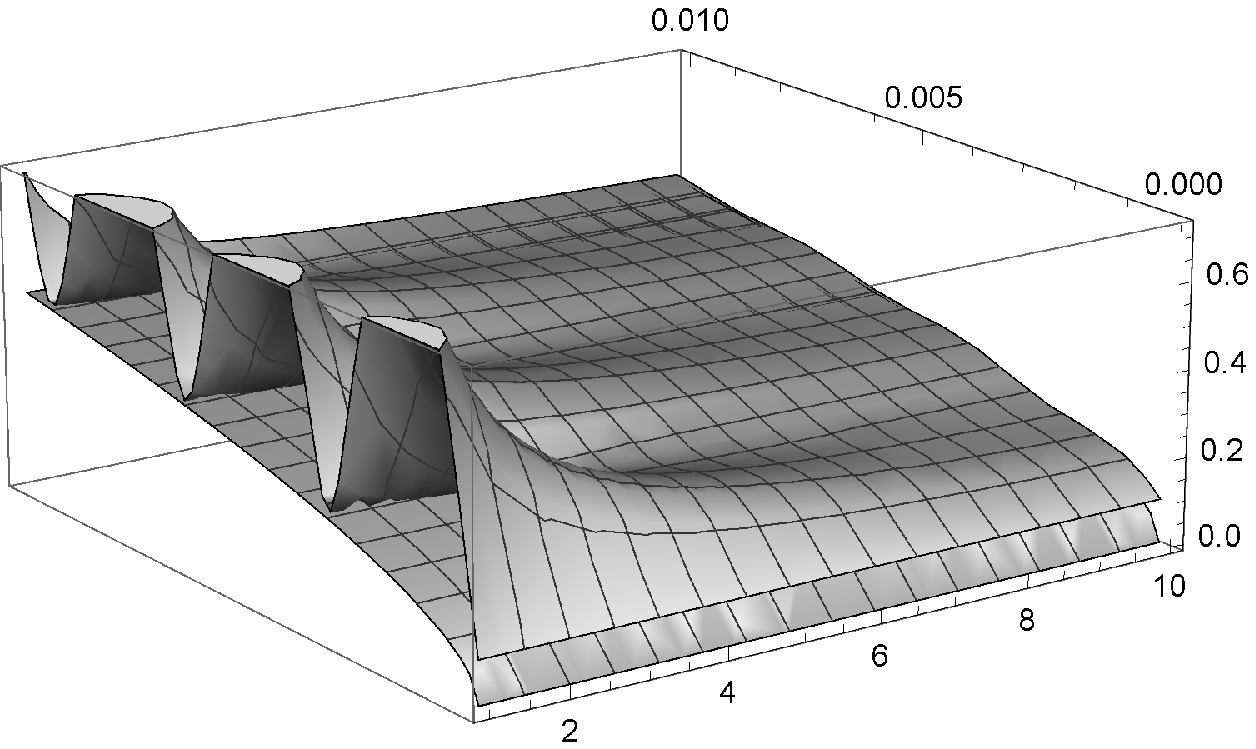}
  \caption{}
\end{subfigure}%
\begin{subfigure}{.5\textwidth}
  \centering
  \includegraphics[width=.8\linewidth]{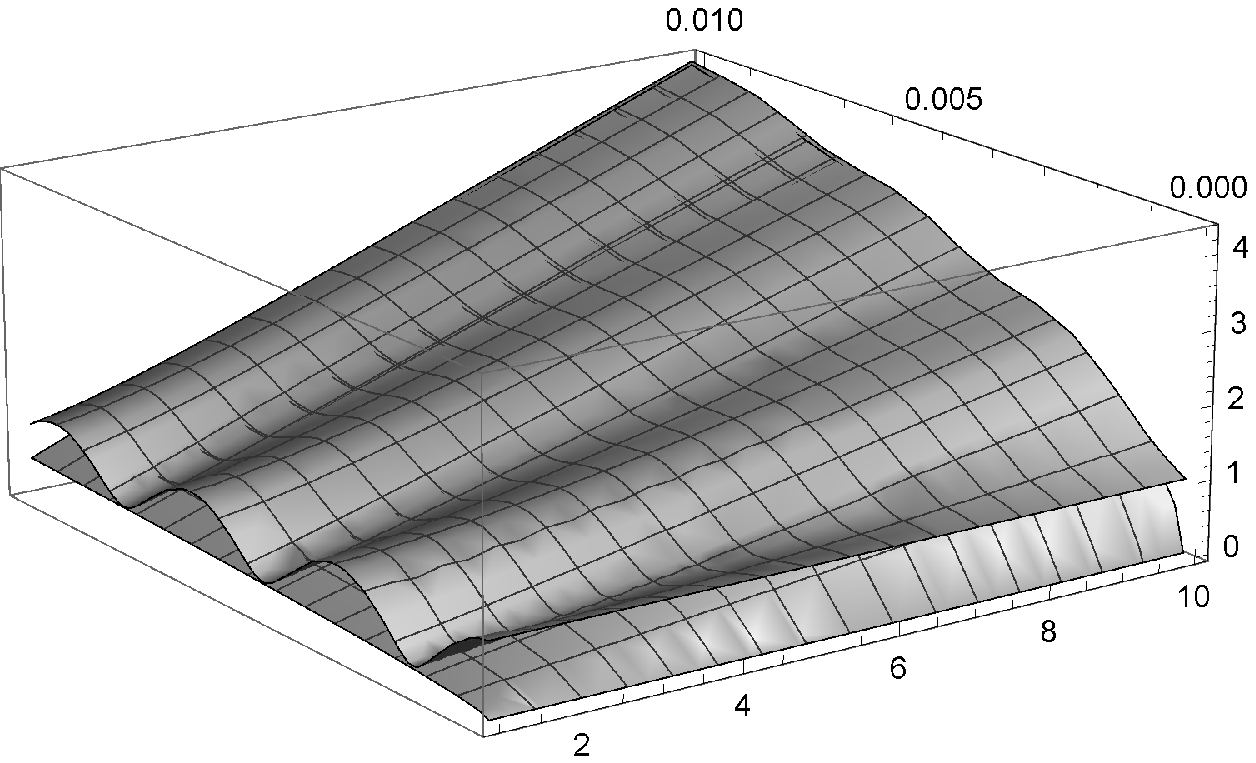}
  \caption{}
\end{subfigure}
\caption{Harmonic rotator, underdamped regime: (a) angle, (b) angular momentum. $\tau\in[0, 0.01]$, $N\in[1, 10]$.} \label{Fig2}
\end{figure}

For the chosen time scale, the transition to the classical dynamics is not as fast as for the case of the free rotator, especially for the less-stable propellers, such as the smaller propellers for $\varphi$ and the larger propellers for $L_z$. Hence the role of the external field that, in principle, may be a control-field for the rotator system. Regarding $\varphi$, the oscillations due to the harmonic potential are larger both in magnitude and the duration for smaller than for the larger propellers. Regarding the angular momentum, initially, the oscillations are larger for larger propellers. The magnitude of change for the angular momentum is smaller than for the case of the free rotator. On average, it's larger than the magnitude of change for the angle observable.

The results regarding the covariance function are qualitatively similar with the results for the angle-observable.

\bigskip

{\bf 3.2 Non-underdamped regime}

\bigskip

Regarding the free rotator, the underdamped and non-underdamped regimes are practically indistinguishable.

\begin{figure}[h]
\begin{subfigure}{.5\textwidth}
  \centering
  \includegraphics[width=.8\linewidth]{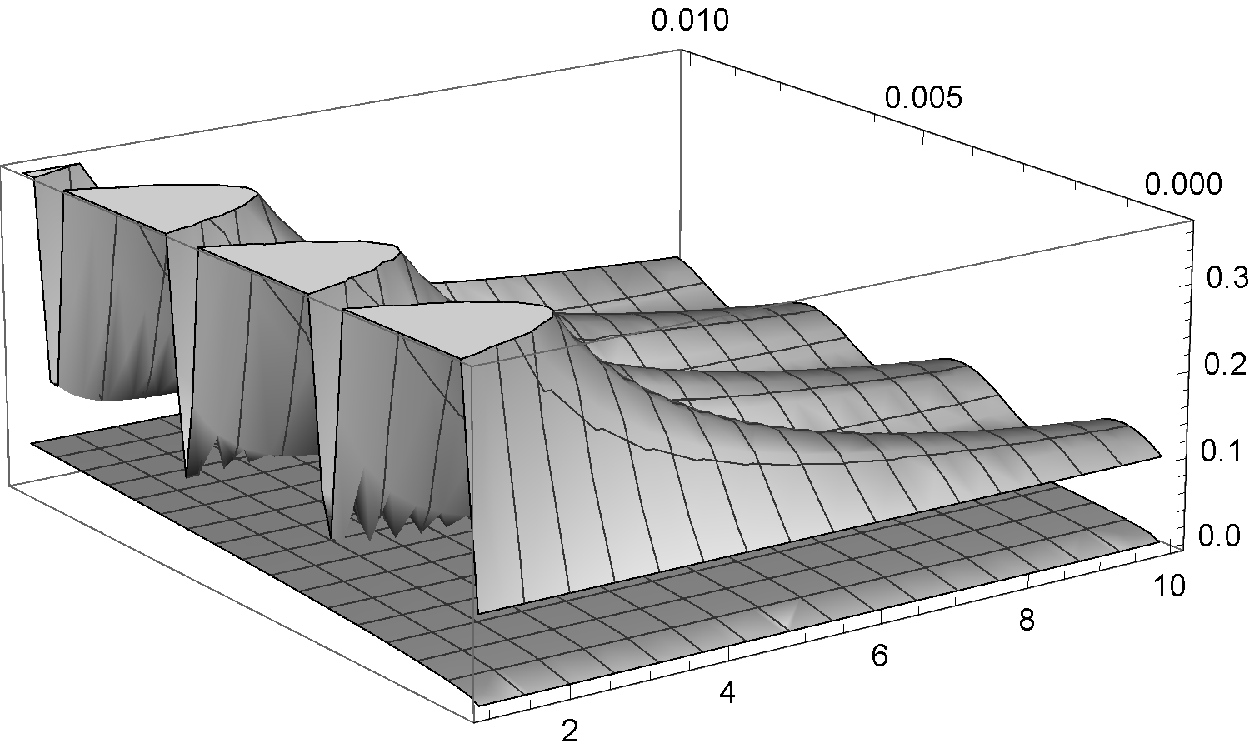}
  \caption{}
\end{subfigure}%
\begin{subfigure}{.5\textwidth}
  \centering
  \includegraphics[width=.8\linewidth]{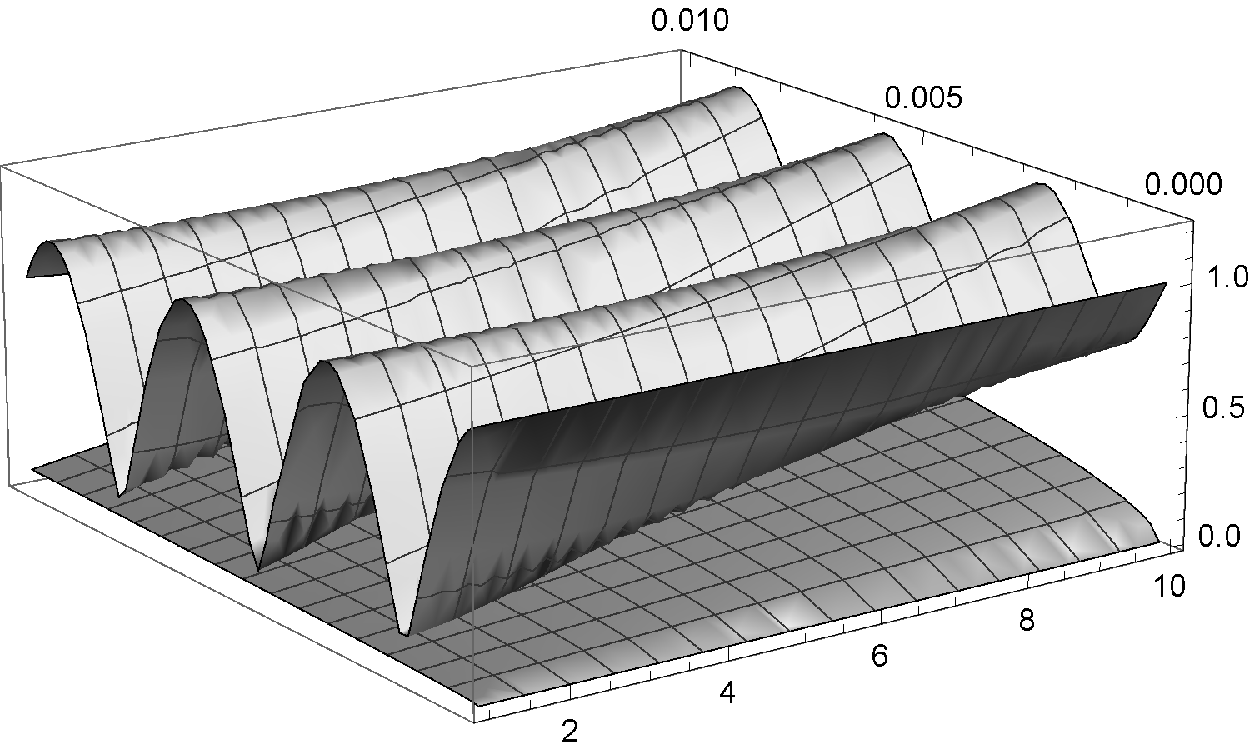}
  \caption{}
\end{subfigure}
\caption{ Harmonic rotator, non-underdamped: (a) angle, (b) angular momentum. $\tau\in[0, 0.01]$, $N\in[1, 10]$.} \label{Fig3}
\end{figure}

From Figure 4 we can learn that, as distinct from the underdamped regime, the classical values exhibit the constant increase, while  the quantum corrections exhibit
the constant decrease with the oscillations; this is obvious for the larger values of $\tau$ that is not presented here. The less stable dynamics is found for the angle-observable for smaller rotators--in contrast with the angular momentum.
Quantum corrections are more pronounced and last longer than for the underdamped regime. The magnitudes of change for the angular momentum are larger than for the angle-observable,
but with the  lower values than for the underdamped regime.

The results regarding the covariance function are qualitatively similar with the results for the angle-observable.

\bigskip

{\bf 4. Discussion}

\bigskip

The Langevin equation for the rotation angle $\varphi$ , cf. e.g. equation (1) in Ref. [6], is a classical-theoretical model and a precursor for aiming at the desired molecular rotators (e.g. cogwheels).
To this end, our starting point, equation (1), can be regarded as a proper quantum mechanical counterpart that assumes the weak-coupling limit and high temperature of the thermal bath as well as the constraint $1\le N \le10$ for the number $N$ of the blades. Weak interaction (small damping factor $\gamma$) and high temperature are the basic assumptions of the {\it microscopic} derivation of the equation (1) that are adopted in Sections 2 and 3.
Relaxing some of those assumptions may be justified for certain situations and also to lead to extensions of our considerations. Nevertheless, this should be performed with caution. Let us briefly emphasize some directions and the possible pitfalls.

A more general model going beyond the model eq.(1) accounts for both reorientation of the rotation axis as well as arbitrary angle of rotation [26]. Adding the "minimally invasive" term to the Caldeira-Leggett equation  leads to a Markovian (Lindblad-form) dynamics of the rotator [27]. Nevertheless, this way the microscopic basis for the equation is abandoned as well as questioned appearance and the microscopic origin of the quantum dissipation term for the thermal bath consisting of the set of mutually independent modes (harmonic oscillators) [28]. The Caldeira-Leggett equation (1) may also be regarded as a {\it phenomenological} equation, in which case the values of both the damping constant $\gamma$ and the bath's temperature $T$ may be considered unrestricted [28].
On the other hand, respecting the microscopic context for arbitrary strength of interaction and temperature is known to be allowed for the restricted set of states--the Gaussian states (cf. Section 4.6 in Ref. [14]).
Therefore, going beyond the small $\gamma$ and/or large $T$ comes at
the price of the questionable microscopic origin or of a restriction to a special set of states of the Brownian particle.

In Fig.1 we assume the blades are connected via the central atom.
For the case of the central disc or a ring  [6], the total moment of
inertia $I = I_{disc} + N I_{\circ} = (\kappa + N) I_{\circ}$, and
analogously for the strength of interaction $\alpha$, thus
numerically reducing this situation to the situations considered in Section 3 while assuming $\kappa < N_{\max}$.
It is worth emphasizing that  the choice of $N_{\max}> 10$ does not qualitatively change our conclusions. Nevertheless,
our restriction to the number of blades satisfying $1\le N\le 10$ is two-fold. First, it aims at the expected realistic chemical molecular species (including the artificially produced ones) [6]. Second, large values of $N$ drive the considerations out of the basic assumptions of our considerations. On the one hand, for some large $N > 10$, according to Fig.1, the blades become close to each other and thus
their environments become rather small, in contrast to the basic assumption of the Caldeira-Leggett model [13, 14]. On the other hand, in the limit $N\to\infty$, a propeller takes the form of a homogeneous disc--which
is a new kind of the rotator's geometry (shape) that requires an independent analysis to be presented elsewhere.

Certain directions of generalization of our starting point eq.(1) exhibit {\it limitations} of our considerations. Assuming non-Markovian environment in the microscopic derivation leads to a generalization of equation (1) that is the exact (and completely positive) master equation for the microscopic model of the quantum Brownian motion [29]. The possibility of the free choice of the strength of interaction and the temperature of the harmonic bath as well as a freedom in the choice of the spectral density go beyond the limitations imposed by the original equation (1) adopted in Section 3 of this paper. On the other hand, going beyond the microscopic modelling opens the door for the more general stochastic processes, such as the L\' evy processes that represent a natural extension of the standard Brownian processes [14, 30]. However,  for certain L\' evy processes, the use of the mean values and standard deviations is not a reliable tool for statistical analysis [14, 30]. Therefore this kind of generalizations is basically complementary to our considerations and requires a separate analysis, see e.g. [31].

Importance of the relation between the open system's geometry (structure) and "function" (dynamics) has already been emphasized in the similar contexts [30-34]. There, the damping coefficient $\gamma$ and the temperature $T$ are recognized to characterize the role of the environmental fluctuations to the open system's dynamics. For the sake of comparison, without further ado, below we give the results that do not presuppose any restrictions of the values of $\gamma$ or $T$ that include also the overdamped regime for the rotator. To this end, in analogy with Section 3, we use the {\it exact} results for $\sigma_{\varphi}$ and $\sigma_L$ that are presented in Appendix II.

The standard deviations generally increase with the increase of the temperature. A slight discrepancy of the classical and quantum results are found only for the angle observable for short time (here not presented); this discrepancy disappears for the longer time intervals and for the larger values of both $\gamma$ and $T$. Below, the results regarding the exact quantum expressions are given. For all figures  we use the following values: $t=0.01, \omega=10, I=1, \sigma_{\varphi}(0) = \sigma(0) =0.1, \sigma_L(0) = 1, \hbar=1$ (in their respective SI units).

Figure 5a depicts the results for the angle observable of the free rotator.
For the harmonic rotator (Fig.5b) for shorter times (t=0.01),
we can see decrease of the standard deviation for lower temperatures like in Fig.5a and, first some increase and then decrease of the standard deviation for larger values of $T$.
 This interplay between the roles of $\gamma$ and $T$, Fig.5b, does not appear for any other (quantum or classical) case.

Figure 6 depicts the results for the angular momentum for both short and longer times. As distinct from the case of the free rotator (Fig.6a), there is a saturation of the standard deviation for the harmonic rotator (Fig.6b)  for larger $\gamma$. Compared to Figure 5, we can see the faster and much larger {\it increase} of $\sigma_L$.

\begin{figure}[h]
\begin{subfigure}{.5\textwidth}
  \centering
  \includegraphics[width=.8\linewidth]{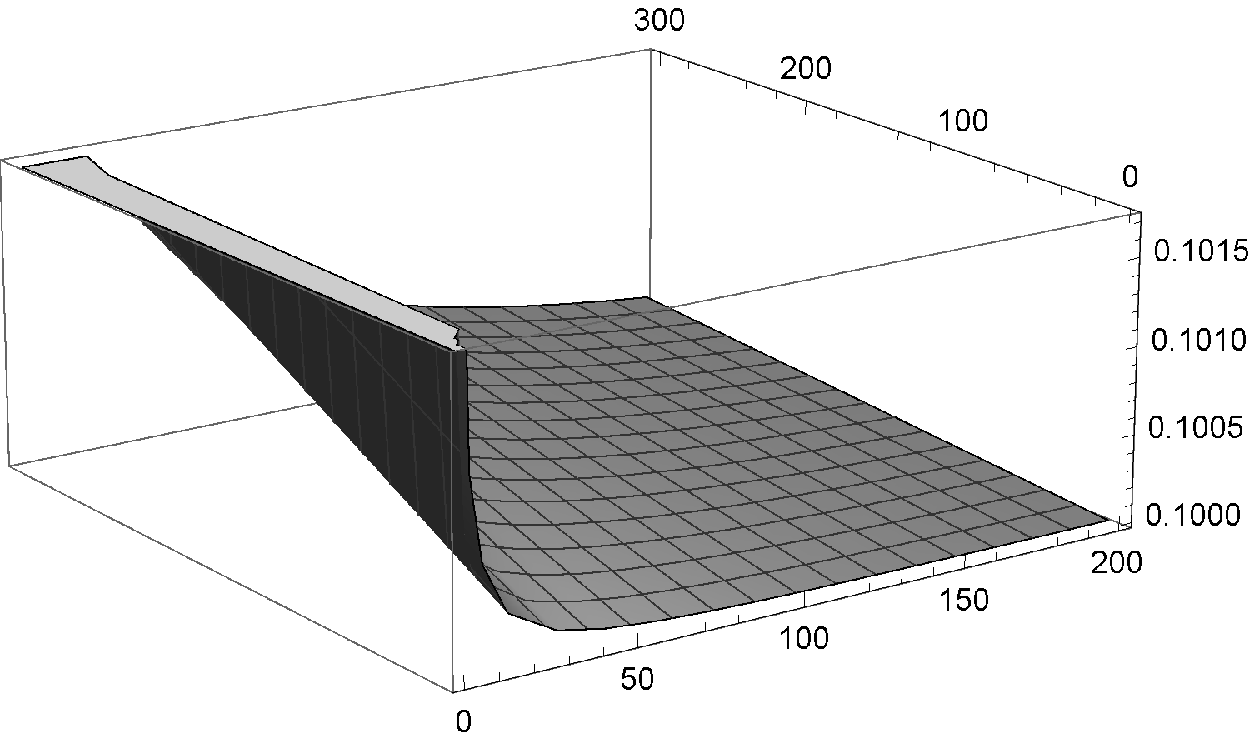}
  \caption{}
\end{subfigure}%
\begin{subfigure}{.5\textwidth}
  \centering
  \includegraphics[width=.8\linewidth]{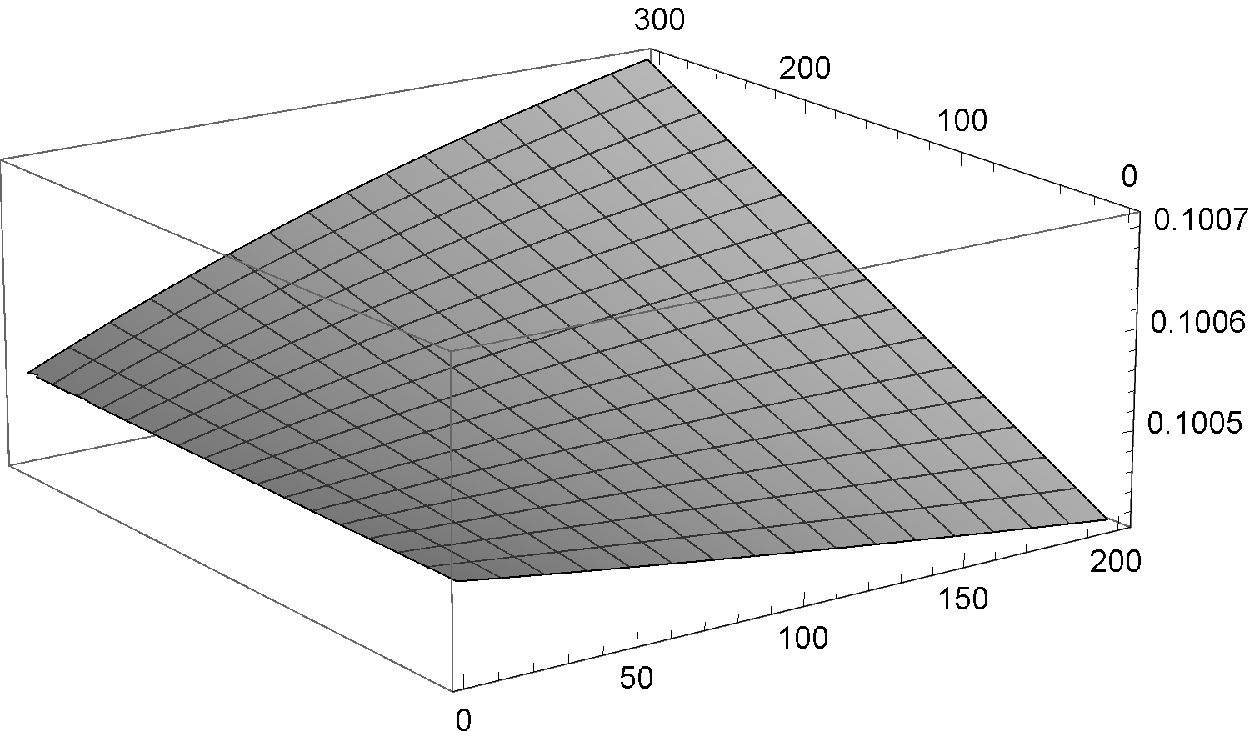}
  \caption{}
\end{subfigure}
\caption{ Dependence of the angle-standard-deviation  on the damping factor, $\gamma\in [1, 200]$ and temperature, $k_BT\in [0.01,300]$. (a) Free rotator, and (b) Rotator in the external harmonic field.} \label{Fig3}
\end{figure}

\begin{figure}[h]
\begin{subfigure}{.5\textwidth}
  \centering
  \includegraphics[width=.8\linewidth]{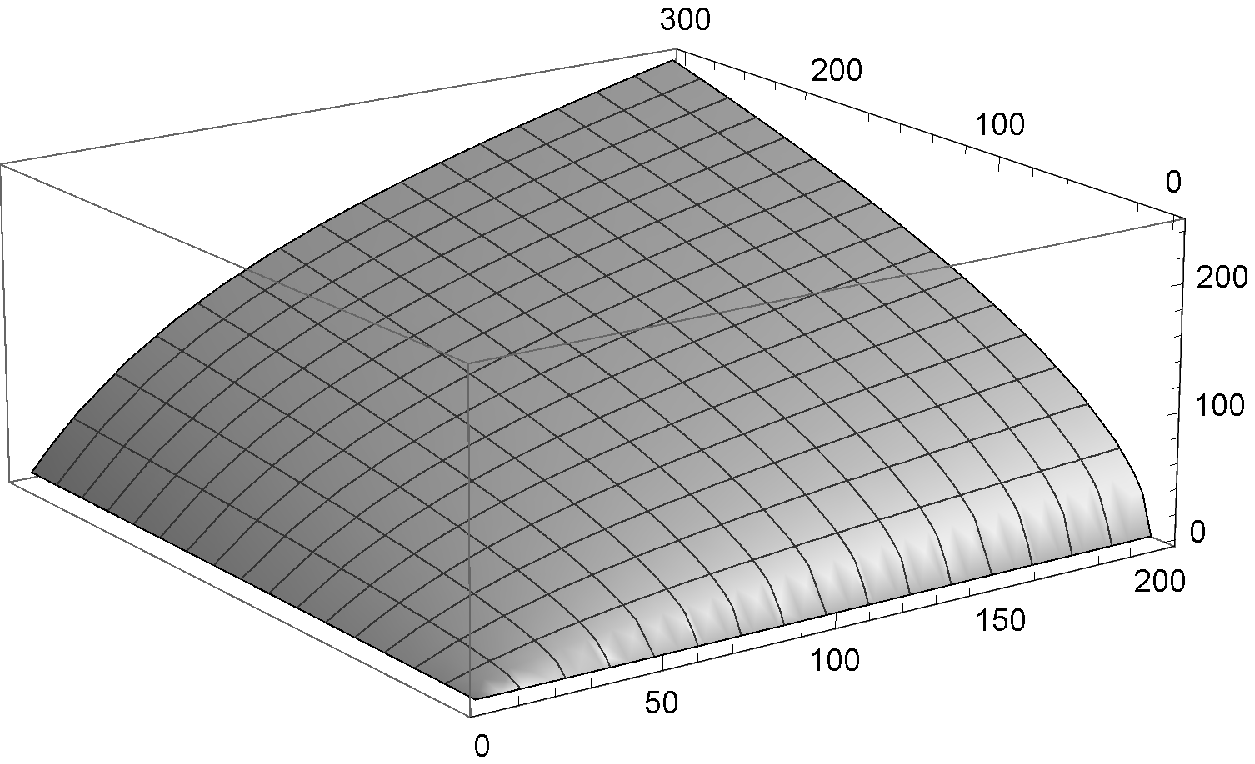}
  \caption{}
\end{subfigure}%
\begin{subfigure}{.5\textwidth}
  \centering
  \includegraphics[width=.8\linewidth]{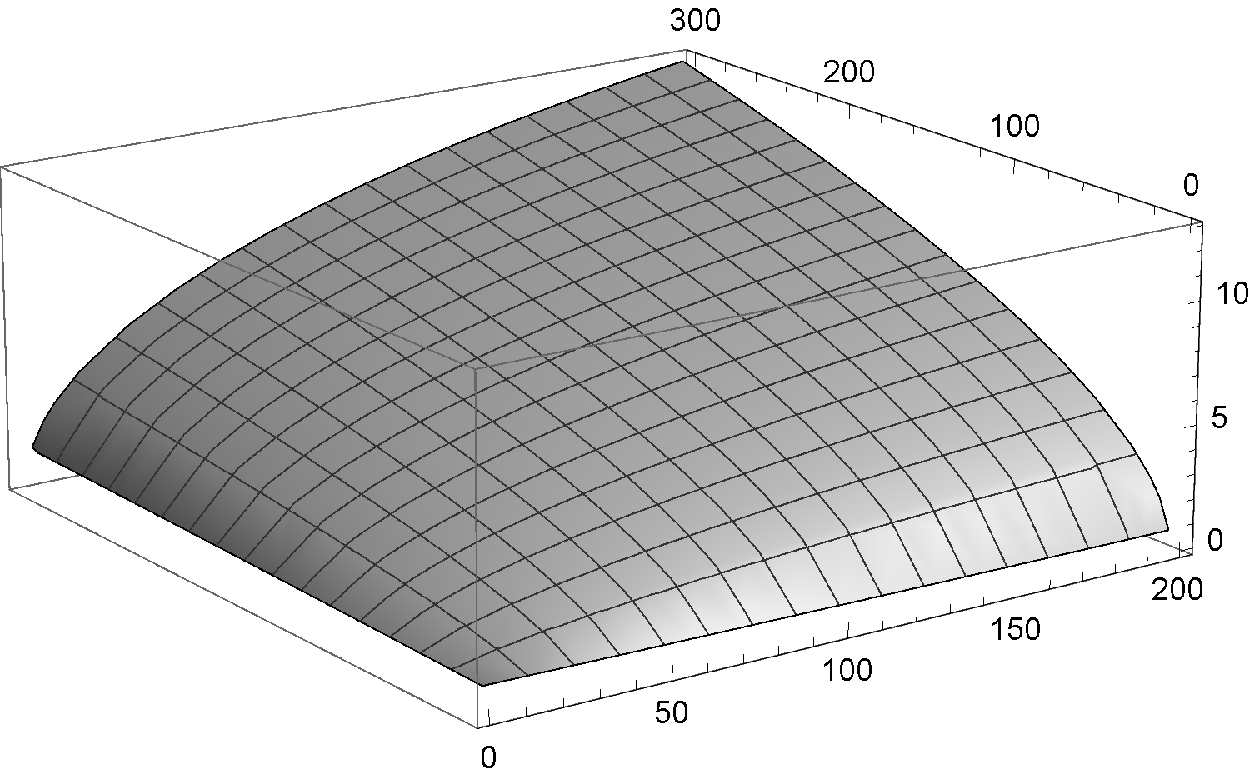}
  \caption{}
\end{subfigure}
\caption{  Dependence of the angular-momentum standard deviation  on the damping factor, $\gamma\in [1, 200]$ and temperature, $k_BT\in [0.01,300]$. (a) Free rotator, and (b) Rotator in the external harmonic field.} \label{Fig3}
\end{figure}

Increase of the temperature leads to the increase of the standard deviations, generally, that is, leads to the decrease of the rotator's stability.
On the other hand, we find monotonic (the absence of the local maximums/minimums) dependence of the standard deviations on $\gamma$ in all cases, except for the case presented in Fig.5b.
For larger $\gamma$, we find better stability (smaller the standard deviation) for the angle observable--in contrast to the angular momentum observable.
Better stability of the angle due to the stronger interaction with the environment may be regarded analogous to the    "noise-enhanced stability (NSE)"  effect [32-34] (and the references therein). That is, the environmental fluctuations may enhance stability of the open system's dynamics [10, 32-34].
The presence of non-monotonic dependence  on $\gamma$ and $T$ in the NSE effect appears for the {\it nonzero, nonquadratic} potentials [32-34] that are beyond the models considered in this paper.

Experience with the {\it isolated} quantum systems may seem to impose the requirements of the rotational symmetry, $A(\varphi) = A(\varphi+2\pi)$, for  relevant quantities of the model, such as those in eq.(2).
However, the absence of the rotational symmetry for the total system's Hamiltonian as well as of the master equation (1) [21] reject the symmetry requirement--typical for the {\it open}
systems, whose symmetries  are open issue in the foundations of the theory. In any case, rotational symmetry is irrelevant for our considerations, which are constrained (Section 2) by the requirements of the small individual rotation  and small initial $\Delta \varphi$.

From Section 3 we  expectably learn about the essentially classical-like dynamics for the time intervals larger than $(N\gamma_{\circ})^{-1}$.
Transition from the exact quantum to the classical-like dynamics goes beyond the decoherence process, which is described by the third term
on the rhs of eq.(1) and is known to be much faster than the processes on the time scale of $(N\gamma_{\circ})^{-1}$ [14, 35].
The "pure decoherence" (also known as the "recoilless") limit of eq.(1) [13, 14]  assumes sufficiently large moment of inertia $I = N I_{\circ}$, for $N\gg 1$, independently of the damping rate $\gamma$. In this limit, [irrespective of the above remarks regarding large $N$], the second (the dissipation) term on the rhs of eq.(1) becomes negligible compared to the third (the decoherence) term.  In Appendix III, we provide the results regarding the "pure decoherence" limit of eq.(1)
that both qualitatively as well as quantitatively departure from the equations (3) and (5), which follow from the exact equation (1), even for the small time intervals $N\tau\ll 1$.
Hence transitions to the classical-like dynamics for eqs.(3) and (5), that are provided in Section 3 and Appendix I, cannot be either reduced to or built solely on the decoherence process.

Therefore we find, that the spatial size and shape matter on the nano-scale even for the simplest possible one-dimensional rotator. That is, transition to the classical dynamics for the realistic {\it finite-size}, $N\sim 10$, rotators (described by the {\it full} eq.(1)) is qualitatively different from the case of (physically unachievable) $N \gg 1$, which regards the pure decoherence limit of eq.(1), even for the small time intervals ($N\tau\ll 1$).

Ever since Hund [36] it is clear that quantum-mechanical nature of molecules does not obviously support the phenomenological existence of a definite
spatial shape of the molecules; consequently, the moment of inertia, $I$, should be a dynamical variable rather than a c-number appearing in eq.(1). In the history of quantum theory and its
applications, different proposals have been used to resolve this riddle, see
e.g. [35] (and the reference therein). Nowadays, it is a common position that
the process of quantum decoherence is responsible for the effective classical
behavior of certain degrees of freedom of open quantum systems [14, 35]. That is precisely the basis of our considerations in which the moment of
inertia is a time-independent real number, not a dynamical degree of freedom
or a variable of the (rigid) rotator system. More precisely, we deal with the
spatial shape and size of a molecule as with a quasi-classical (environment-induced) characteristic [37], which provides a basis of the treatment of the moment of inertia
as a real, time-independent classical parameter for a rigid rotator as appearing
in eqs.(1) and (2) as well as in the results presented in this paper. A microscopic model that would describe
the shape-inducing decoherence [35, 37] as well as decoherence described by eq.(1) or regarded in Refs. [26, 27] might significantly improve our
understanding of the emergence of classical from the quantum level [35].

As distinct from the classical counterparts [6-9, 17], the molecular cogwheels suffer from the quantum mechanical corrections for short time intervals; those corrections make the rotation even less stable (less predictable) than in the classical case. While individually small (and monotonically decreasing in time), those corrections, for every number $N$ of the blades, may uncontrollably {\it accumulate} due to a {\it series of repeated external actions}, which  are not taken into account by equations (3) and (5). That is, unless some kind of "error correction" is performed, the final value of a standard deviation for one action becomes the initial value for the next one and hence accumulation of the quantum-mechanical contributions is unavoidable. That is, a series of classical actions (e.g. of $40-100$ "kicks" [38])
may uncontrollably   increase  the quantum contributions  and in principle give rise to the relatively large final values of the standard deviations. Hence a departure from, i.e. a limited use of, the standard semi-classical treatment [6-12, 39]--that is valid on the time scales larger than $(N\gamma_{\circ})^{-1}$--where the quantum contributions are effectively lost.

The results of Section 3  reveal the absence of the simple rules for utilizing the stability and the desired control of the molecular Brownian rotators. That is, there is not such thing as conditions for "the most stable rotation". Instead, "relative stability" appears as a combination of a number of criteria stemming from Section 3. We may say that this reminds us of the  reasoning typical for the engineering optimization methods [16]--already at the microscopic scale. Below, we emphasize and briefly comment  those criteria, all of them stemming from  Section 3.

(A) The choice of the observable to be acted on.

\noindent {\it Comments}. Application of the external electric field on the polar molecules exerts an action on the rotator's angle, while for the magnetic molecules,
application of the magnetic field exerts an action on the molecule's angular-momentum. The actions may be practically free choice only in the deep-classical domain, which is defined in the point (C) below.
The choice is additionally laden by the point (B) below as well as by the observation that the angle and the angular momentum exhibit the opposite dependence on the size $N$ of the rotator.

(B) The magnitude of change of the standard deviations.

\noindent {\it Comments}. The magnitudes of change of the standard deviations for $\varphi$ and $L_z$ differ by several orders  for the free rotator, and for about one order  for the harmonic rotator.
Those magnitudes are larger for the angular momentum than  for the angle of rotation.
Nevertheless, the quantitative details are sensitive to the size of the rotator as well as of the criteria below described in (C) and (D).

(C) The rate of the external operations and the time-scale of "relaxation" of rotation.

\noindent {\it Comments}. The "relaxation" times in eqs.(3) and (5) are of the order of $(N\gamma_{\circ})^{-1}$. For the time intervals much longer than $(N\gamma_{\circ})^{-1}$, the rotator dynamics is practically
indistinguishable from the classically predicted dynamics. However, for the time-intervals shorter than $(N\gamma_{\circ})^{-1}$, the rotator dynamics is subject of the quantum-mechanical corrections, which, albeit individually negligible, may still be of interest in certain situations.
For the damping rate $\gamma_{\circ} \le 10^8$ (e.g. for the value $1.3\cdot 10^6$Hz for the nonpolar solvent as the  environment for the internal rotation regarding the toluene molecule [21]), the typical rates of operation of the order of  $10^{-12}$s [40-43] {\it deeply fall  in the quantum mechanical domain}. That is, such operations are by $10^3$ (i.e. by $10^5$) times faster than the relaxation rate $(10\gamma_{\circ})^{-1}$. Then a series of the external actions described above, in the time-window  shorter than $(10\gamma_{\circ})^{-1}$, may lead to the
non-negligible increase of the quantum contributions and therefore to less predictable (less stable) rotation dynamics.
 The choice of the faster/slower operations is a matter of both the desired functioning of the molecular cogwheels as well as of the available techniques. Quantitatively, the domain (quantum or classical) of the behavior depends on the regime as emphasized in the point (E) below.

(D) The presence/absence of the external field(s).

\noindent {\it Comments}. Free rotator exhibits fast transition to the classical dynamics. Rotator in the external harmonic field is subject of fast oscillations in the quantum domain. Observability of these oscillations depends on both the rate of the external actions as well as on the time-resolutions of the available measurement techniques. E.g., if the measurement cannot resolve between the adjacent maxima/minima, the results will generally provide the time-averaged data, even in the quantum domain, which is more pronounced for the non-underdamped regime.

(E) The underdamped/non-underdamped regime.

\noindent {\it Comments}. Generally, the two regimes are  similar. Quantitatively though, the non-underdamped regime emphasizes the longer-lasting quantum behavior for the harmonic rotator. Thus we can expect non-trivial dependence on the criteria (A)-(D) in conjunction with the presence/absence of the external field.

Designing the proper scenarios for utilizing the relative stability of the molecular nano-rotators is beyond the scope of the present paper. To this end, the work is in progress as well as accounting for the non-quadratic "internal" and time-dependent driving fields for the rotator system [6, 30-34].  The results will be presented elsewhere.

\bigskip

{\bf 5. Conclusion}

\bigskip

Molecular cogwheels as the functional parts of the desired nano-machines are with the definite geometrical shape and the finite spatial size. If modelled as the one-dimensional, propeller-shaped rigid rotators,
their dynamical behavior exhibits  dependence on the spatial size in regard of the role of quantum decoherence, the possible accumulation of the quantum-mechanical contributions as well as on devising  the optimal scenarios for utilizing the relative stability of rotation.

\bigskip

{\bf Acknowledgements} The work on this paper is partially financially supported by Ministry of education, science and technological development, Serbia, grant no 171028.

\bigskip

{\bf Appendix I. The classical expressions for the first and second moments}

\bigskip

Classical expressions for  the equations (3) and (5) follow from these equations after substituting the classically allowed null initial
values, $\sigma_{\varphi}(0) = \sigma_L(0) = \sigma(0) = 0$. The proof of this claim  is as follows.

All expressions for the classical translational variables [2, 3] are equally valid, {\it mutatis mutandis}, for the classical rotational variables [6].
Bearing in mind that the classical Langevin equation is the classical limit of eq.(1) [21],
we simply exchange the translational by the rotational variables in the expressions known for the classical translational variables; in our notation: $f=2I\gamma$ in eq.(9).

Eq.(10) in [2] gives the mean square displacement for the classical
Brownian particle:

\begin{equation}
\langle (x(t) - x_0)^2\rangle = {2mk_BT\over f^2} \left({f\over
m}t - 1 + e^{-ft/m} \right).
\end{equation}

\noindent where $x_0\equiv x(0)$.

It is easy to express the standard deviation for the angle
$\varphi$:

\begin{equation}
(\Delta \varphi(t))^2 = \langle (\varphi(t) - \langle
\varphi(t)\rangle)^2\rangle = \langle (\varphi(t) -
\varphi_0)^2\rangle- (\langle\varphi(t)\rangle -  \varphi_0)^2.
\end{equation}

In our notation, eq.(22) of [2] reads:

\begin{equation}
\langle\varphi(t)\rangle -  \varphi_0 = {L_z(0)\over 2I\gamma}
\left(1 - e^{-2\gamma t} \right).
\end{equation}

Now placing eq.(9) and eq.(11) into eq.(10) with the use of
{\it equipartition}, $L_z^2(0) = Ik_BT$:

\begin{equation}
(\Delta \varphi(t))^2 =  {k_B T \over I\gamma^2} [\gamma t - (1 -
e^{-2\gamma t}) + {1\over 4}(1 - e^{-4\gamma t})].
\end{equation}

On the other hand, equations (12) and (14) from [2] in our notation read:

\begin{equation}
\langle L_z(t)\rangle = \langle L_z(0)\rangle e^{-2\gamma t},
\langle L_z^2(t)\rangle = Ik_BT + \left( \langle L_z(0)^2\rangle -
Ik_BT\right) e^{-4\gamma t}.
\end{equation}

From eq.(13) easily follows:

\begin{equation}  (\Delta L_z(t))^2 = \langle L_z^2(t)\rangle - \langle L_z(t)\rangle^2 = Ik_B T (1 -
e^{-4\gamma t} ).
\end{equation}

Equations (12) and (14) can be easily recognized in eq.(3), where the quantum terms are those proportional to
$\sigma_{\varphi}(0), \sigma_L(0)$ and $\sigma(0)$.

Finally for the free rotator, we derive the covariance $\sigma_{\varphi L}$. From Section II of Ref. [2]:

\begin{equation}
{d\varphi\over dt} = {L_z\over I}, \quad {dL_z\over dt} = -\beta L_z + I A(t),
\end{equation}

\noindent with the zero average for the stochastic force, $I\langle A(t)\rangle = 0$. Then

\begin{equation}
{d\over dt}(\langle \varphi L_z\rangle - \langle\varphi\rangle \langle L_z\rangle) =  \langle {d\varphi\over dt}L_z + \varphi{dL_z\over dt}\rangle -
\langle{d\varphi\over dt}\rangle \langle L_z\rangle - \langle{dL_z\over dt}\rangle \langle\varphi\rangle
\end{equation}

Substituting eq.(15) into eq.(16), we obtain the differential equation for $\sigma_{\varphi L}$:

\begin{equation}
{d\sigma_{\varphi L}\over dt} = {(\Delta L_z)^2\over I} - \beta \sigma_{\varphi L},
\end{equation}

\noindent whose integration, for $\beta = 2\gamma$, gives the expression eq.(3) for $\sigma_{\varphi L}$ (in eq.(3) we use the dimensionless $\sigma$).

For the harmonic rotation, we directly take over the classical expressions, equation (55)
in [3]:

\begin{eqnarray}
&\nonumber& (\Delta\varphi(t))^2 = {D\over \beta\omega^2} \left(1
- {e^{-\beta t}\over \omega^2_1} (\omega^2_1 + {\beta^2\over 2}
\sin^2\omega_1t + \beta\omega_1\sin\omega_1\cos\omega_1t)\right)
\\&&\nonumber
(\Delta L_z(t))^2 = {I^2D\over \beta} \left(1 - {e^{-\beta t}\over
\omega^2_1} (\omega^2_1 - {\beta^2\over 2} \sin^2\omega_1t -
\beta\omega_1\sin\omega_1\cos\omega_1t)\right)\\&&
\sigma_{\varphi L} = {D\over \omega_1^2} e^{-\beta t} \sin^2\omega_1 t,
\end{eqnarray}

\noindent where $\omega_1=\sqrt{\omega^2-\beta^2/4}$ and $D= \beta
k_BT/I$; in our notation,  $\beta=2\gamma$. Within the
approximation of the order $O(\gamma/\omega)$ that is used in
eq.(5), all the sine and cosine terms in eq.(18) disappear while
$\omega_1\approx \omega$. The remaining classical terms
are easy recognized in eq.(5), in which the quantum terms are proportional to
$\sigma_{\varphi}(0), \sigma_L(0)$ and $\sigma(0)$.

\bigskip

{\bf Appendix II. Exact results for the harmonic rotator}

\bigskip

Solutions of eq.(2) are presented by eq.(5) in the approximate form, assuming weak interaction (i.e. $\gamma_{\circ}/\omega \ll 1$).
In this appendix we provide the exact solutions of eq.(2) that underlie eq.(5) and are not explicitly given in Ref.[25], where the method of calculation is presented in detail; we use $\Omega^2 = \gamma^2 - \omega^2$.

\begin{eqnarray}
&\nonumber&
(\Delta\varphi(t))^2 = {k_BT\over I\omega^2\Omega^2}(\Omega^2 +e^{-2\gamma t}
(\omega^2 - \gamma^2 \cosh(2\Omega t) -\gamma\Omega\sinh(2\Omega t)))+
\\&&\nonumber
{(\Delta L_z^2(0))^2\over I^2\Omega^2} e^{-2\gamma t} \sinh^2(\Omega t) +{(\Delta\varphi(0))^2\over \Omega^2}e^{-2\gamma t}(-\omega^2 \cosh^2(\Omega t)+
\\&&\nonumber
\gamma^2\cosh(2\Omega t)+\gamma\Omega \sinh(2\Omega t)) + {e^{-2\gamma t}\sigma_{L\varphi}(0)\over2I\Omega^2}(2\gamma \sinh^2(\Omega t)+
\\&&
\Omega\sinh(2\Omega t)),
\end{eqnarray}

\begin{eqnarray}
&\nonumber&
(\Delta L_z(t))^2 = {Ik_BT\over \Omega^2}(-\omega^2(1-e^{-2\gamma t}) + \gamma^2(1 - e^{-2\gamma t}\cosh(2\Omega t)) -
\\&&\nonumber
\gamma\Omega e^{-2\gamma t}\sinh(2\Omega t)) + {e^{-2\gamma t}(\Delta L_z(0))^2\over \Omega^2} (-\omega^2\cosh^2(\Omega t)+\gamma^2\cosh(2\Omega t) -
\\&&\nonumber
\gamma\Omega\sinh(2\Omega t)) + {e^{-2\gamma t}I^2\omega^4 (\Delta\varphi(0))^2\over \Omega^2}\sinh^2(\Omega t)+
\\&&
{e^{-2\gamma t}I\omega^2 \sigma_{L\varphi}(0)\over2\Omega^2}(
2\gamma\sinh^2(\Omega t)-\Omega\sinh(2\Omega t)),
\end{eqnarray}

\begin{eqnarray}
&\nonumber&
\sigma_{L\varphi}(t) = e^{-2\gamma t}({4\gamma k_BT\over \Omega^2}
\sinh^2(\Omega t)  + {(\Delta L_z(0))^2\over I\Omega^2}  (-2\gamma\sinh^2(\Omega t)+\Omega\sinh(2\Omega t))-
\\&&
{I\omega^2\over\Omega^2} (\Delta\varphi(0))^2(2\gamma\sinh^2(\Omega t) + \Omega\sinh(2\Omega t))
- {\sigma_{L\varphi}(0)\over\Omega^2} (\omega^2\cosh(2\Omega t) - \gamma^2))
\end{eqnarray}

Taking the approximation $\gamma_{\circ}/\omega\ll 1$ (when $\Omega^2\approx - \omega^2$ and $\Omega \approx \imath\omega$) and, for simplicity, $\langle L_z(0)\rangle = 0 = \langle\varphi(0)\rangle$,  eqs.(19)-(21) lead to eq.(5). Equations (19)-(21) are used for Figures 5 and 6 in the main text.

\bigskip

{\bf Appendix III. The decoherence limit}

\bigskip

The decoherence limit of eq.(1) follows for the sufficiently large moment of inertia, $I=NI_{\circ}, N\gg 1$. Then the second (the dissipation) term in eq.(1) can be neglected compared to the third (the decoherence term); the ratio of the third and the second term is proportional to $N$. Therefore, for simplicity, we place $\gamma = \gamma_{\circ}$ and $I = N I_{\circ}$.
Then instead of eq.(2), we obtain:

\begin{eqnarray}
&\nonumber& {d\langle\varphi(t)\rangle\over dt} = {1\over I}
\langle L_z(t)\rangle,\\&& {d\langle L_z(t)\rangle\over dt} = -
\langle V'(\varphi(t))\rangle,
\nonumber
\\&&
{d\langle \varphi^2(t)\rangle\over dt} = {1\over I} \langle
L_z(t)\varphi(t) + \varphi(t) L_z(t) \rangle \nonumber,
\\&& \nonumber
{d\langle L_z^2(t)\rangle\over dt} = - \langle L_z(t)
V'(\varphi(t)) + V'(\varphi(t)) L_z(t)\rangle  + 4I\gamma k_B T
\\&& {d\over dt}\langle \varphi L_z + L_z\varphi \rangle = {2\over
I}\langle L_z^2\rangle -2\langle \varphi V'(\varphi)\rangle.
\end{eqnarray}

From eq.(22) it is straightforward to obtain:

\begin{eqnarray}
&\nonumber&
\sigma_{\varphi}^2 = \sigma_{\varphi}^2(0)+ {\sigma(0) \over N} \tau +  {\sigma_L^2(0) \over N^2} \tau^2 + {4k_BT\over 3NI_{\circ}\gamma_{\circ}^2}\tau^3,
\\&&
\sigma_L^2 = \sigma_L^2(0) + {4Nk_BT\over I_{\circ}\gamma_{\circ}^2} \tau,
\end{eqnarray}

\noindent for the free rotator, and the exact expressions:

\begin{eqnarray}
&\nonumber&
\sigma_{\varphi}^2 = \sigma_{\varphi}^2(0) \cos^2{\omega \over\gamma_{\circ}} \tau
+  {\sigma^2_L(0) \over N^2} \sin^2 {\omega\over\gamma_{\circ}} \tau +
{\sigma(0) \over 2N} \sin 2{\omega\over\gamma_{\circ}} \tau\\&&\nonumber  + {2 k_BT\over NI_{\circ}\omega^2} \tau - {\gamma_{\circ} k_BT\over NI_{\circ} \omega^3} \sin 2{\omega\over\gamma_{\circ}} \tau,\\&&
\sigma_L^2 = {\sigma_L^2(0)}\cos^2 {\omega\over\gamma_{\circ}} \tau + N^2\sigma_{\varphi}^2(0)\sin^2 {\omega\over\gamma_{\circ}} \tau - {N\sigma(0)\over 2}  \sin 2{\omega\over\gamma_{\circ}} \tau\\&&\nonumber + {2N k_BT\over I_{\circ}\omega^2} \tau + {N\gamma_{\circ} k_BT\over I_{\circ} \omega^3} \sin 2{\omega\over\gamma_{\circ}} \tau
\end{eqnarray}

\noindent for the harmonic rotator. In Section 3 it is assumed that $\gamma_{\circ}/\omega\sim 10^{-3}$.

Expression for $\sigma_{\varphi}^2$ in eq.(23) is of the form of eq.(3.442) in Ref. [14], which is obtained for the initial wave-packet with $\sigma(0)=0$ for {\it short} time intervals, i.e. for $\tau\ll 1$. The rest of eqs.(23) and (24) cannot be obtained as approximations of the  equations (3) and (5). Even for the short time intervals allowing for the approximation $e^{-N\tau}\approx 1 - N\tau$, the two descriptions departure from each other. It is rather obvious that equations (23) and (24) give different predictions
than equations (3) and (5) for the finite $\tau$  as well as in the asymptotic limit $\tau\to\infty$.
Therefore we can conclude that the exact dynamics eqs.(3) and (5) as well as the related transitions to the classical behavior--that are presented in Section 3 and in Appendix I--are {\it not} determined by decoherence for the {\it finite-dimensional} ($N\le 10$) rotators, even for the small time intervals $N\tau\ll 1$.

\bigskip

{\bf References}

\bigskip

[1] Einstein A 1905 {\it Ann. d. Physik} 1V, 549

[2] Uhlenbeck G E and Ornstein L S 1930 {\it Phys. Rev.} {\bf 36} 823

[3] Wang M C and Uhlenbeck G E 1945 {\it Rev. Mod. Phys.} {\bf 17} 323

[4] Chandrasekhar S 1993 {\it Rev. Mod. Phys.} {\bf 15} 1

[5] ten Hagen B, van Teeffelen S and L\" owen H 2011 {\it J. Phys.: Condens. Matter} {\bf 23} 194119

[6] Kottas G S, Clarke L I, Horinek D, Michl J  2005 {\it J. Chem. Rev.} {\bf 105}
1281

[7] Goodsell D S 2008 {\it The Machinery of Life} (Copernicus, Springer Verlag: New
York, NY).

[8] Browne W R and Feringa B L 2006 {\it Nature Nanotech.} {\bf 1} 25

[9] Coffey W T, Kalmykov Y P and Titov S V 2004 {\it The Journal of Chemical Physics} {\bf 120} 9199

[10] Riemann P 2002 {\it Phys. Rep.} {\bf 361} 57

[11] ten Hagen B, van Teeffelen S and L\" owen H 2011 {\it J. Phys.: Condens. Matter} {\bf 23} 194119

[12] Reimann P and  H\" anggi P 1998 {\it Chaos} {\bf 8} 629

[13] Caldeira A O and Leggett A J 1983 {\it Physica A: Stat. Mech. Applicat.} {\bf 121}
587

[14] Breuer H-P and Petruccione F 2002 {\it The theory of open quantum systems} (Clarendon Press, Oxford, NY)

[15] Meng X, Zhang J-W and Guo H 2016 {\it Physica A} {\bf 452} 281

[16] Ravindran A, Ragsdell K M, and Reklaitis G V 2006 {\it Engineering Optimization: Methods and Applications} (2nd Edition, Wiley)

[17] Rusconi C C, P\" ochhacker V, Kustura K, Cirac J I and  Romero-Isart O 2017 {\it Phys. Rev. Lett.} {\bf 119} 167202

[18] Shigley E J and Mischke R C 2001 {\it Mechanical Engineering Design} (Mc Graw Hill, International Edition)

[19] Rondin L, Gieseler J,   Ricci F, Quidant R,   Dellago C and Novotny L 2017 {\it Direct Measurement of Kramers’ Turnover with a Levitated Nanoparticle}
arXiv:1703.07699v2 [cond-mat.mes-hall]

[20] Rivas \' A and  Huelga S F 2011 {\it Open quantum systems. An introduction} (SpringerBriefs in Physics, Springer, Berlin)

[21] Suzuki Y and Tanimura Y 2001 {\it J. Phys. Soc. Japan} {\bf 70} 1167

[22] Breitenberger E 1985 {\it Found. Phys.} {\bf 15} 353

[23] Jordan P 1927 {\it Z. Phys.} {\bf 44} 1

[24] Deck R T and  \" Ozt\" urk N 1994 {\it Found. Phys. Lett.} {\bf 7} 419

[25] Petrovi\' c I and  Jekni\' c-Dugi\' c J 2018 {\it Facta Universitatis} in press

[26] Stickler B A, Schrinski B, and Hornberger K 2017 Rotational friction and diffusion of quantum rotors
E-print archive arXiv: 1712.05163v1 [quant-ph]

[27] Papendell B, Stickler B A, and Hornberger K 2017 Quantum angular momentum diffusion of rigid
bodies, E-print archive arXiv:1712.05596v1 [quant-ph]

[28] Ferialdi L 2017  	{\it Phys. Rev. A} {\bf 95} 052109

[29] Hu B L, Paz J P, and  Zhang Y 1992 {\it Phys Rev. D} {\bf 45} 2843

[30]  Lisowski B, Valenti D, Spagnolo B, Bier M, and  Gudowska-Nowak E 2015 {\it Phys. Rev. E}
{\bf 91} 042713

[31] Valenti D, Guarcello C, and Spagnolo B (2014) {\it Phys. Rev. B} {\bf 89} 214510

[32] Fiasconaro A, Mazo J J, and Spagnolo B 2010 {\it Phys. Rev. E} {\bf 82} 041120

[33] Agudov N V,  Dubkova A A, and Spagnolo B (2003) {\it Physica A} {\bf 325} 144

[34] Valenti D, Magazzu L,  Caldara O, and  Spagnolo B (2915) {\it Phys. Rev. B} {\bf 91} 235412

[35] Giulini D, Joos E,  Kiefer C, Kupsch J, Stamatescu I-O and  Zeh H D 1996
{\it Decoherence and the Appearance of a Classical World in Quantum Theory}
(Springer: Berlin)

[36] Hund F 1927 {\it Z. Phys.} {\bf 43} 805

[37] Jekni\' c-Dugi\' c J 2009 {\it Eur. Phys. J.} D {\bf 51} 193

[38]  Gadway B, Reeves J, Krinner L and Schneble D 2013 {\it Phys. Rev. Lett.} {\bf 110} 190401

[39] Vacek J and  Michl J 2001 {\it Proc. Nat. Acad. Sci.}  {\bf 98} 5481

[40] Lin K, Song Q, Gong X,  Ji Q,  Pan H, Ding J, Zeng H  and Wu J 2016 {\it Phys. Rev. A} {\bf 92} 013410

[41] Sch\" onborn J B, Herges R and Hartke B 2009  {\it J. Chem. Phys.} {\bf 130} 234906

[42]  Hoki K, Yamaki M and  Fujimura Y 2003 {\it Angew. Chem.} {\bf 115} 3083

[43]  Chocholousova J, Vacek J, Kobr L and Miller J 2006 {\it in Proceeding HPCMP-UGC '06
Proceedings of the HPCMP Users Group Conference, IEEE
Computer Society Washington, DC, USA}, pp. 26-29

\end{document}